\input{aipcheck}

\documentclass[
    ,final            
  ]
  {aipproc}

\layoutstyle{8x11single}
\setcitestyle{numbers}

\begin{document}

\title{  $f_0(600)$, $\kappa(800)$, $\rho(770)$ and
$K^*(892)$, quark mass dependence from unitarized  SU(3) Chiral 
Perturbation Theory}


\author{J. Nebreda}{
 }

\author{J.R. Pel\'aez}{
 address={Dept. F\'isica Te\'orica II. Universidad Complutense, 28040, Madrid. Spain}
 }

\begin{abstract}
We study the strange and non-strange quark mass dependence 
of the parameters of the $f_0(600)$, $\kappa(800)$, $\rho(770)$ and
$K^*(892)$ resonances generated from
elastic meson-meson scattering using unitarized one-loop 
Chiral Perturbation Theory. 
We fit simultaneously all experimental
scattering data up to 0.8-1 GeV together with 
lattice  results on decay constants and scattering lengths up to a pion mass of
440 MeV. 
Then, the strange and non-strange quark masses are varied 
from the chiral limit up to values of interest for
lattice studies. In these amplitudes,  the mass and width of the $\rho(770)$ and
$K^*(892)$ present
a similar and smooth quark mass dependence.
In contrast, both
scalars present a similar non-analyticity at high quark masses. Nevertheless
the $f_0(600)$ dependence on both quark masses is stronger
 than for the $\kappa(800)$ and the vectors.
We also confirm the lattice assumption of quark mass independence
of the vector two-meson coupling that, in contrast,
is violated for scalars.  
\end{abstract}

\maketitle


\section{Introduction}
Although QCD is well established as the theory of strong interactions, 
the hadronic realm is beyond the reach of perturbative calculations.
In that regime, lattice methods are a useful tool to
calculate QCD observables, but results on light  meson
resonances are few and usually 
obtained at very large quark masses compared with their
physical values. 
Very recently \cite{Hanhart:2008mx}, 
an alternative technique, based on 
Chiral Perturbation Theory (ChPT) and dispersion relations,
has been applied to calculate the 
dependence of the $f_0(600)$ (or ``sigma'') and
$\rho(770)$ resonances on the averaged u and d quark mass, $\hat{m}$. 
In this talk we report our progress extending this study to include the strange quark
within a unitarized SU(3) ChPT formalism \cite{Nebreda}. Our aim is threefold:
to confirm previous results within a more general formalism, to analyze the  dependence on the 
$\hat m$
of the $K^*(892)$ and
$\kappa(800)$ strange resonances and then, 
to study the dependence of all the
$f_0(600)$, $\kappa(800)$, $\rho(770)$ and
$K^*(892)$ parameters in terms of the strange quark mass $m_s$.

For unitarization we use the well-known one-loop elastic 
Inverse Amplitude Method (IAM) that provides a remarkable description of
$\pi\pi$ and $K\pi$ data up to $\sim 1$ GeV, while generating the
poles associated to the $f_0(600)$, $\rho(770)$, $K^*(892)$
and $\kappa(800)$ resonances \cite{Dobado:1996ps}. 
This is achieved using Low Energy
Constants (LECs)
compatible with those of standard ChPT.
Previous descriptions come from fitting 
only to experiment (see details in \cite{GomezNicola:2001as}),
and therefore are mostly sensitive to the LECs 
that govern the $s$ dependence of partial waves.
In order to get better determinations of the LECs that 
carry an explicit meson mass
dependence, now we have fitted also to lattice results on $M_\pi$, $M_K$,
$f_\pi$, $f_K$ and scattering lenghts \cite{lattice}.

\begin{figure}
\begin{tabular}{c}
  \includegraphics[scale=.45,angle=-90]{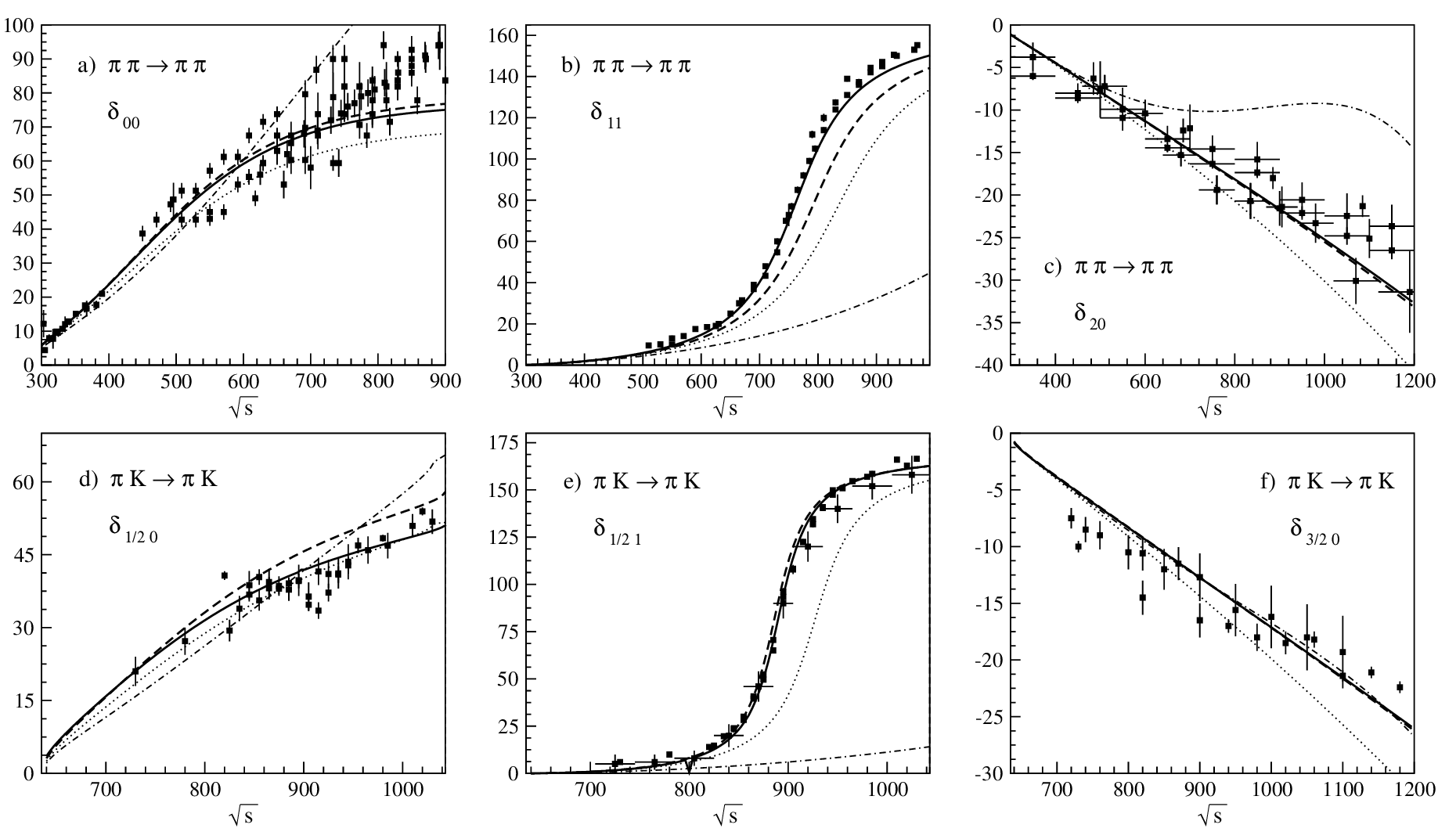}\\
  \includegraphics[scale=.45,angle=-90]{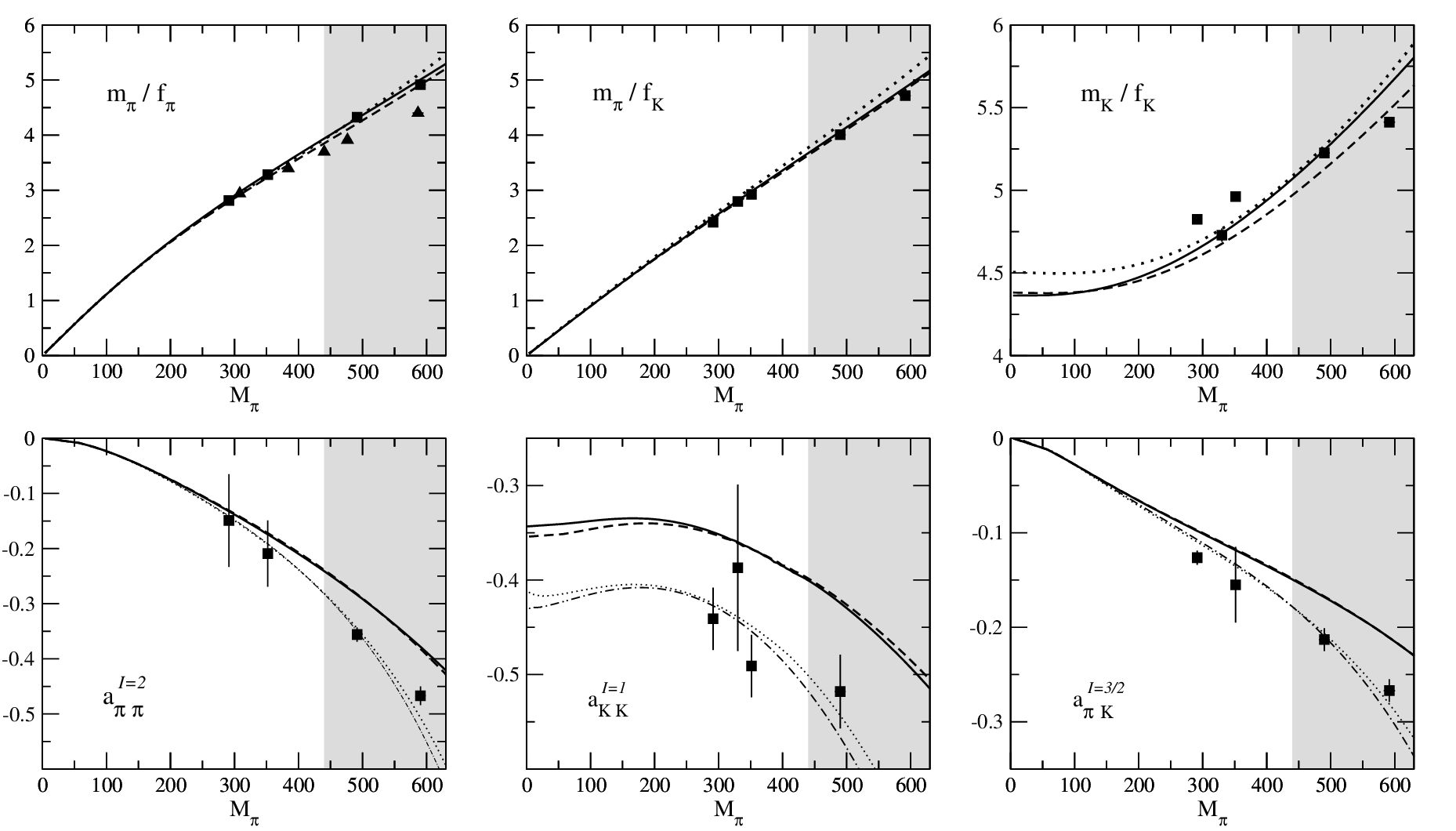}
  \end{tabular}
  \caption{
 Two upper rows: results of our IAM fits versus experimental data on $\pi\pi$ and $\pi K$
scattering. Two lower rows: result of the unitarized fits to lattice calculations of $f_\pi$, $f_K$,
$m_\pi/f_\pi$ and the $\pi^+\pi^+$, $K^+ K^+$, $K^+\pi^+$ 
scattering lengths.
The continuous and dashed lines correspond, respectively, to Fits I and II. For comparison
we show the results of the IAM 
if we used the ChPT LECs obtained from $K_{l4}$ \cite{Amoros:2001cp} (dotted line) and the results of
standard non-unitarized ChPT with the sets of LECs given in \cite{Pelaez:2004xp} (dot-dashed line). The lattice data come from \cite{lattice} and the references for the experimental data are given in \cite{GomezNicola:2001as}.
}
  \label{fig:fits}
\end{figure}

In Fig.~\ref{fig:fits} we present two new fits: "Fit I" describes best
the data, but we show also "Fit II" to give an idea of the typical size 
of the systematic uncertainties. It also makes clear that we 
don't need any fine tuning of the LECs 
to describe the experimental and lattice results together.

\vspace*{-.4cm}
\section{Dependence on the mass of the light quarks} 

The $\hat m$ is a parameter of ChPT closely related to 
the pion mass that we can vary to study the corresponding
change in the resonances.
It is known \cite{Dobado:1996ps} that the IAM works for Goldstone bosons masses 
at least as high as 500 MeV. Since we want pions 
always lighter than kaons and etas in order to apply the elastic 
approximation, we will show results up to 
$M_\pi< 440\,\rm MeV$ but not beyond, since then 
$M_K\simeq 600\,\rm MeV$.
In terms of quark masses,  this means $\hat m/\hat m_{\rm phys}\leq 9$.

\begin{figure}
\begin{tabular}{c}
  \includegraphics[scale=.53]{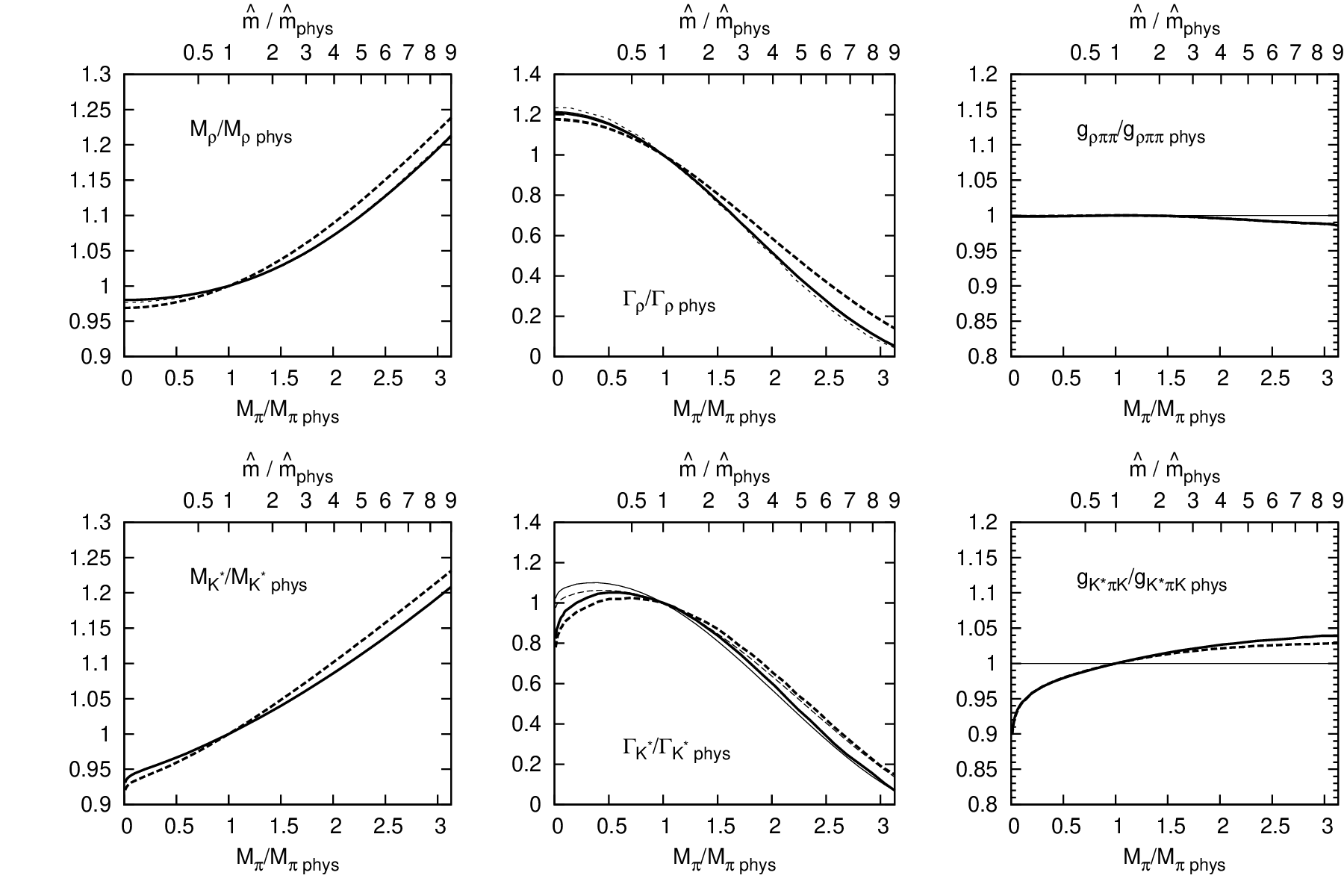}\\
  \includegraphics[scale=.53]{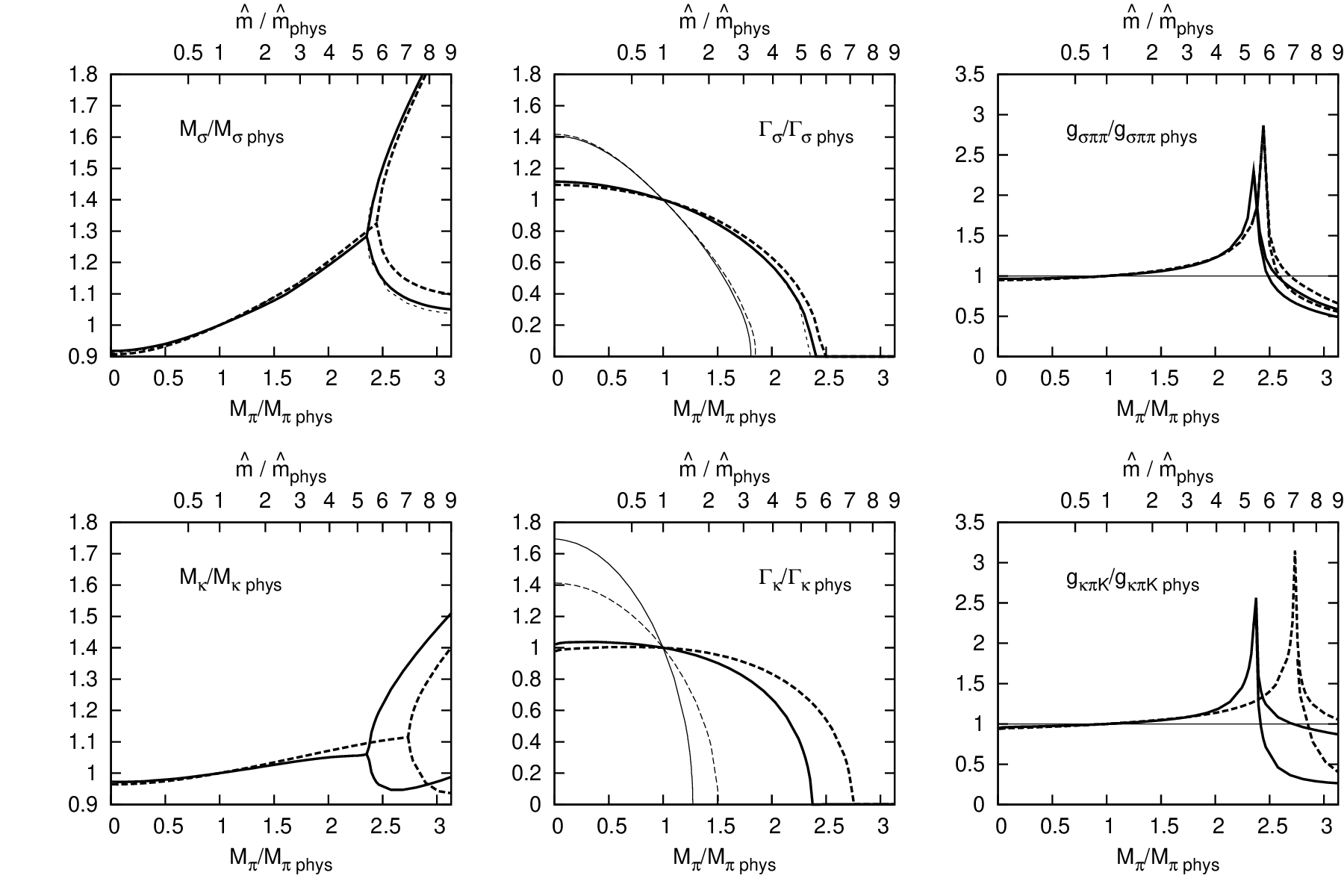}
  \caption{sigmakappa}
  \end{tabular}
  \label{fig:mhatdependence}
  \caption{Dependence of the $\rho(770)$, $K^*(892)$, $f_0(600)$ 
 and $\kappa(800)$ mass, width and coupling to two mesons with respect to
the non-strange quark mass $\hat m$  (horizontal upper scale), or
the pion mass (horizontal lower scale).
Note that we give all quantities normalized to their 
physical values. 
The thick continuous and dashed lines correspond to
Fit I and Fit II, respectively.
For the $\rho(770)$ and the  $f_0(600)$ these results are very 
compatible with those in \cite{Hanhart:2008mx}
using  SU(2) ChPT (dotted line).
The continuous (dashed) thin line shows the $M_\pi$ dependence of
the widths from the change of phase space only, assuming a constant
coupling of the resonances to two mesons, $\rho(770)$ and $f_0(600)$ 
to $\pi\pi$ and 
$K^*(892)$ and $\kappa(800)$ to $\pi K$, calculated from the
dependence of masses and momenta given by Fit I (II).
}
\end{figure}

\vspace*{-.3cm}
\paragraph{Light vector mesons}
The $\rho(770)$ and $K^*(892)$ vector
resonances are well established $q \bar q$ states belonging to 
an SU(3) octet. In the first two rows of Fig.~\ref{fig:mhatdependence}, we show their dependence 
on the non-strange quark masses. For each resonance, 
its mass and width are defined
from the position of the associated pole in the second Riemann sheet,
through the usual Breit-Wigner identification $\sqrt{s_{pole}}\equiv M-i\,\Gamma/2$, and 
its coupling to two mesons is given by the residue of the amplitude at the pole position.

The results obtained for the $\rho(700)$ resonance are very consistent
with the previous SU(2) results in \cite{Hanhart:2008mx} (dotted line in Fig.~\ref{fig:mhatdependence}), and the estimations 
for the two first coefficients of the $M_\rho$ chiral expansion \cite{bruns}.. Besides, 
the similarity of the behavior of $\rho(770)$ and $K^*(892)$ is evident. 
Their masses increase smoothly 
as the quark mass increases, but much slower than the pion mass.
This means that
 the thresholds grow faster than the masses of the resonances, and
as a consequence there is a strong phase space suppression. 
In fact, the decrease of their widths agrees remarkably well with
the expected reduction coming only from phase space suppression without a dynamical effect
through the vector coupling to two mesons (thin lines). 
Accordingly, we see that $g_{\rho\pi\pi}$ and
$g_{K^*\pi K}$ are remarkably constant,
which is an assumption made in lattice studies
of the $\rho(770)$ width \cite{Aoki:2007rd}.

\vspace*{-.3cm}
\paragraph{Light scalar mesons}
The $f_0(600)$, or sigma, and the $\kappa(800)$ scalar mesons 
are still somewhat controversial. Their huge width makes their experimental identification complicated and there are no present lattice calculations
with realistic quark masses.
It is therefore more interesting to obtain predictions on their 
quark mass dependence. 

In the third and fourth rows of Fig.~\ref{fig:mhatdependence} 
we show their dependece on the non-strange quark mass.
As before, we find that for the $f_0(600)$ the results are 
in good agreement with the existing SU(2) calculation of 
\cite{Hanhart:2008mx}.

The most prominent feature of the scalars behavior
is the appearance of two branches for the mass.
The reason is that for physical values of the quark mass,
the  poles associated to resonances appear as
conjugated poles in the second Riemann sheet. As the quark mass increases these poles 
move closer to the real axis 
until they join in a single pole below threshold and then
split again, now remaining in the real axis.

Despite the evident qualitative similarities, the quantitative behavior of the $f_0(600)$ 
and the kappa is rather different.
In particular, the growth of the $f_0(600)$ mass before the ``splitting point'' is 
much faster than that of the $\kappa(800)$.

Regarding their width decrease, we show that for the scalars it 
cannot be attributed to the phase space reduction
due to the increase of pion and kaon masses.
Related to this, we see that their coupling constants to two mesons show a strong quark mass dependence. Moreover, they 
increase dramatically near the ``apparent splitting point''.

\vspace*{-.2cm}
\section{Dependence on the mass of the strange quark} 

We will only vary the strange quark mass
in the limited range $0.8<m_s/m_{s\,\rm phys}<1.4$ to ensure that the kaon does not become
too heavy to spoil the ChPT convergence nor too light to require a coupled channel
formalism

\begin{figure}
\begin{tabular}{c}
  \includegraphics[scale=.53]{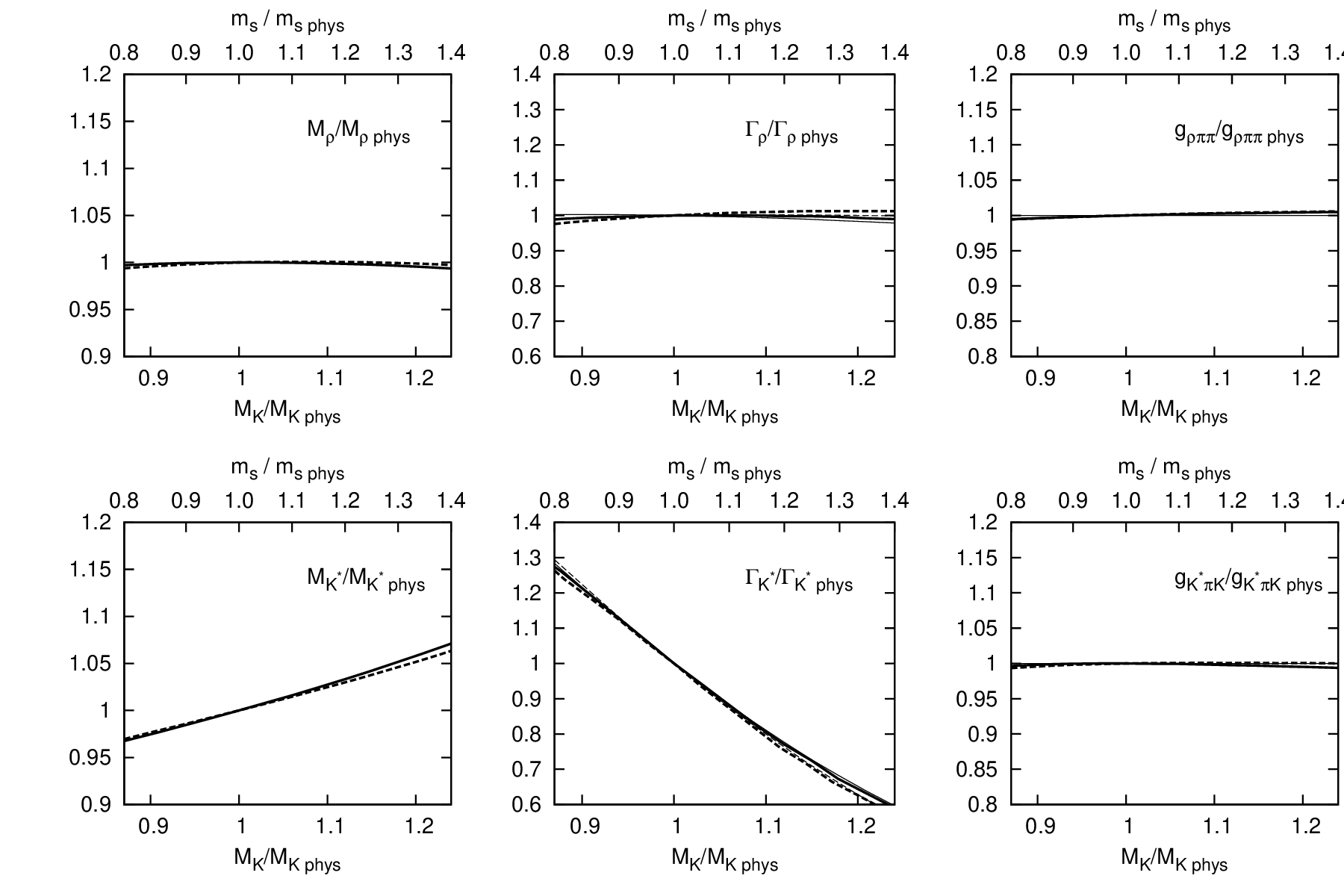}\\
  \includegraphics[scale=.53]{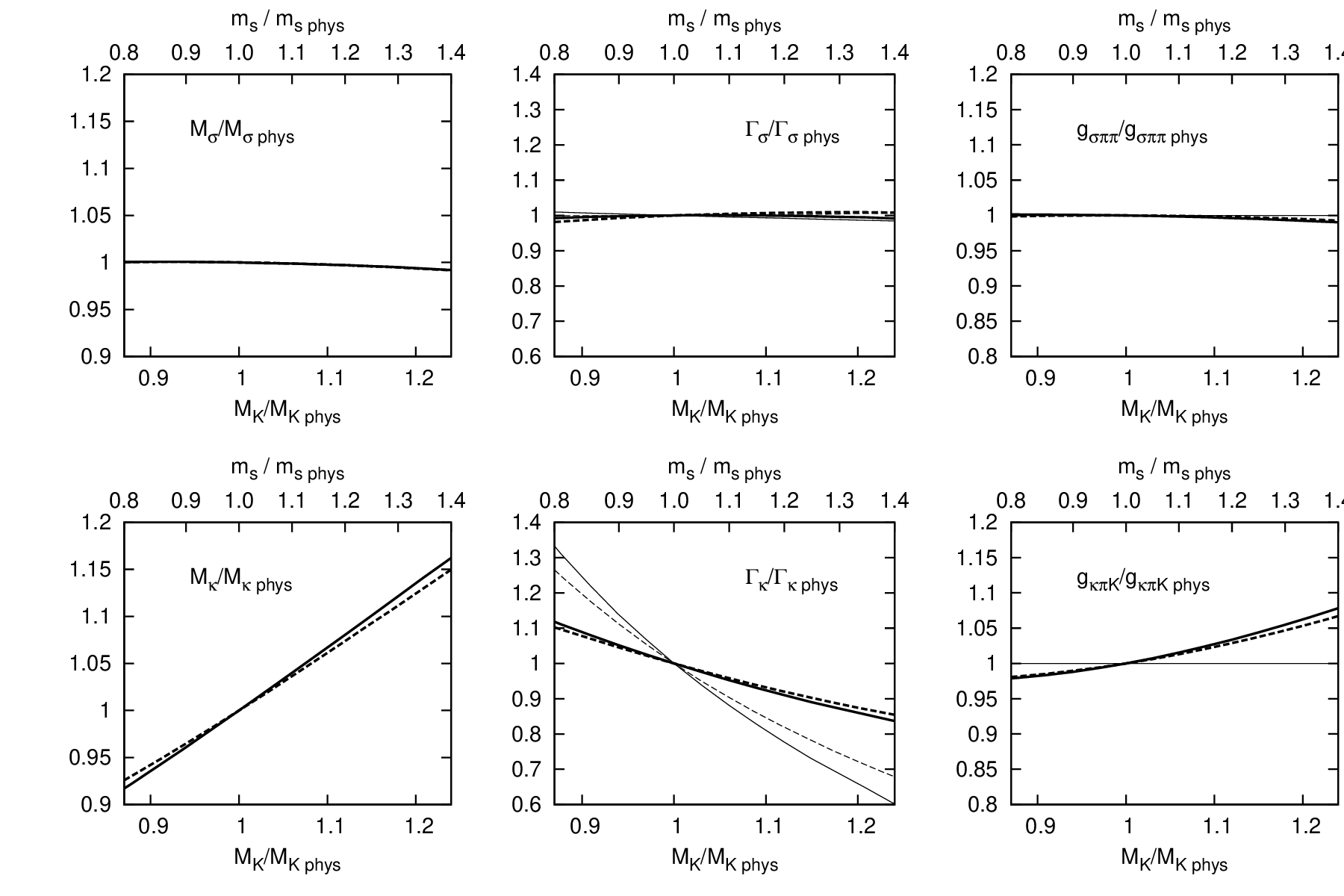}
  \caption{sigmakappa}
  \end{tabular}
  \label{fig:msdependence}
    \caption{Dependence of the $\rho(770)$, $K^*(892)$, $f_0(600)$ 
 and $\kappa(800)$ mass, width and coupling to two mesons with respect to
the strange quark mass $\hat m$  (horizontal upper scale), or
the kaon mass (horizontal lower scale).
Note that we give all quantities normalized to their 
physical values. 
The thick continuous and dashed lines correspond to
Fit I and Fit II, respectively.
The continuous (dashed) thin line shows the $M_K$ dependence of
the widths from the change of phase space only, assuming a constant
coupling of the resonances to two mesons, $\rho(770)$ and $f_0(600)$ 
to $\pi\pi$ and 
$K^*(892)$ and $\kappa(800)$ to $\pi K$, calculated from the
dependence of masses and momenta given by Fit I (II).
}

\end{figure}

\vspace*{-.3cm}
\paragraph{Light vector mesons}
In the two upper rows of Fig.~\ref{fig:msdependence} 
we show the kaon (or strange quark) mass dependence 
of the $\rho$ and $K^*(892)$.
As it could be expected, the properties of the $\rho(770)$ being non-strange 
are almost independent of the strange quark mass within the range of study. 

Obviously,  the $K^* (892)$ shows a strong $m_s$ dependence. 
As the kaon mass is made heavier, the $K^*(892)$ mass grows much
faster than it did when increasing the light quark mass.
On the other hand, the $K^*(892)$ mass grows much slower than the kaon mass,
so that phase space shrinks and the resonance width
decreases almost exactly as it would be expected from phase space suppression only.
Accordingly, its coupling to $K\pi$ is almost constant.

\vspace*{-.3cm}
\paragraph{Light scalar mesons}
In the lower two rows of Fig.~\ref{fig:msdependence}, we show the variation of the sigma 
and $\kappa(800)$ properties.
Again, the effect on the resonance in the $\pi\pi$ chanel, the sigma, is very small.

On the contrary, the mass and width of the $\kappa(800)$
change by as much as 16\% within the range of study. Finally we see again 
that the width decrease deviates significantly from what expected from only phase space 
suppression, in agreement with $g_{\kappa\pi  K}$ depending quite strongly on the strange 
quark mass.

\vspace*{-.3cm}
\section*{Acknowledgments}

\vspace*{-.1cm}
Work partially supported by Spanish Ministerio de
Educaci\'on y Ciencia contracts: FPA2007-29115-E,
FPA2008-00592 and FIS2006-03438,
U.Complutense/Banco Santander grant PR34/07-15875-BSCH and
UCM-BSCH GR58/08 910309 as well as the European Community-Research Infrastructure
Integrating Activity
``Study of Strongly Interacting Matter''
(HadronPhysics2, Grant Agreement
n. 227431)
under the Seventh Framework Programme of the EU.

\end{document}